\documentclass[onecollarge]{svjour2-mod}       
\usepackage[T1]{fontenc}
\usepackage[utf8]{inputenc}

\usepackage{graphicx}
\usepackage{amssymb}
\usepackage{bbm}
\usepackage{xcolor}
\usepackage{wasysym}
\def\defeq{\;\buildrel\hbox{\small def}\over{\,=\;}}

\begin{document}
\title{Ellipsoidal equilibrium figure and Cassini states of rotating planets and satellites deformed by a tidal potential in the spatial case.}
\subtitle{}
\author{Hugo A. Folonier \and Gwena\"el Boué \and Sylvio Ferraz-Mello}

\institute{Hugo A. Folonier \and Sylvio Ferraz-Mello \at Instituto de Astronomia Geofísica e Ci\^{e}ncias Atmosf\'{e}ricas, Universidade de S\~{a}o Paulo, 05508-090 S\~{a}o Paulo, Brazil. \and Hugo A. Folonier \and Gwena\"el Boué \at ASD/IMCCE, CNRS-UMR8028, Observatoire de Paris, PSL, Sorbonne Universit\'{e}, 77 Avenue Denfert-Rochereau, F-75014 Paris, France.}

\titlerunning{}

\maketitle

\begin{abstract}

The equilibrium figure of an inviscid tidally deformed body is the starting point for the construction of many tidal theories such as Darwinian tidal theories or the hydrodynamical Creep tide theory. This paper presents the ellipsoidal equilibrium figure when the spin rate vector of the deformed body is not perpendicular to the plane of motion of the companion. We obtain the equatorial and the polar flattenings as functions of the Jeans and the Maclaurin flattenings, and of the angle $\theta$ between the spin rate vector and the radius vector. The equatorial vertex of the equilibrium ellipsoid does not point toward the companion, which produces a torque perpendicular to the rotation vector, which introduces terms of precession and nutation. We find that the direction of spin may differ significantly from the direction of the principal axis of inertia $C$, so the classical approximation $\tens{I}\vec{\omega} \approx C\vec{\omega}$ only makes sense in the neighborhood of the planar problem. We also study the so-called Cassini states. Neglecting the short-period terms in the differential equation for the spin direction and assuming a uniform precession of the line of the orbital ascending node, we obtain the same differential equation as that found by Colombo (1966). That is, a tidally deformed inviscid body has exactly the same Cassini states as a rotating axisymmetric rigid body, the tidal bulge having no secular effect at first order.

\end{abstract}

\section{Introduction}{\label{sec1}}

The figure of equilibrium of an inviscid tidally deformed body is the basis of many tidal theories, where the viscosity is introduced to the model at a later stage (e.g., Darwin, 1880; Kaula, 1964; Mignard, 1979; Efroimsky and Lainey 2007; Ferraz-Mello et al., 2008; Ferraz-Mello, 2013; Folonier et al., 2018). How the viscosity effect is included is not unique and depends on the tidal approach. 

The problem of finding the equilibrium figure of a self-gravitational rotating body has been extensively studied. If tidal and rotational deformations on an inviscid fluid are small enough, the reference shape can be approximated by different ellipsoids, depending on the case (Chandrasekhar, 1969). When the deformation is solely due to the rotation, the equilibrium figure is an oblate spheroid called ellipsoid of \textit{Maclaurin}. In the case where the deformation is only due to the tide generated by a close-in companion, the equilibrium figure is a prolate ellipsoid called ellipsoid of \textit{Jeans}. Finally, when the deformation is due simultaneously to both rotation and tides, and the direction of rotation is perpendicular to the plane of motion of the companion when the deformation is due only to the tides, the equilibrium figure is a triaxial ellipsoid called ellipsoid of \textit{Roche}. It is important to note that ellipsoidal figures are excellent first approximations, but they are not exact equilibrium figures (Poincaré 1902; Lyapounov 1925, 1927).

In this work, we show that in the case in which the axis of rotation of an extended body is not necessarily perpendicular to the plane of motion of the companion, the body adopts, to the first order in flattenings, a triaxial ellipsoidal shape that differs from the Roche ellipsoids. In addition, we show that the vertex cannot point to the companion, except for the instant in which it passes through the line of the nodes along its orbit. In this way, is obtained the instantaneous orientation of the principal axes of inertia, missing information in most previous studies. In modern tidal theories like Boué et al. (2016), which is a generalization to the spatial case of the tidal theory presented by Correia et al. (2014) for a body with a Maxwell rheology, and Ragazzo and Ruiz (2015) and (2017), the reference ellipsoidal equilibrium figure is implicit in the calculation of the gravitational field, since, according to Love theory, the gravitational potential and the shape are proportional to each other.

On the other hand, in the seventeenth century, Cassini announced its three famous laws for the motion and rotation of the Moon: (1) the spin rate equals the mean orbit rate; (2) the spin pole $\vec{s}$ maintains a constant inclination to the ecliptic pole $\vec{k}$; and (3) the spin pole $\vec{s}$, the orbit pole $\vec{n}$, and the ecliptic pole $\vec{k}$, all remain coplanar (see Cassini, 1693; Tisserand, 1891). Colombo (1966), using a very simple theory applied to an extended axially symmetric body, showed that the synchronous spin-orbit resonance of the first Cassini's law is independent of the secular spin-orbit resonance described by the two last. These results were generalized by Peale (1969) for a triaxial rigid body in any $p:1$ spin-orbit resonance, where $p$ is a half-integer. In this paper, we study Cassini's laws for the particular case of a fluid body without viscosity, where the coefficient $J_2$ (or equivalently the ratio $(C-A)/C$) is not only given by the rotation through Maclaurin flattening but also by the tide given through Jeans flattening.

The layout of the paper is as follows: We first proceed, in Sects. \ref{sec2} and \ref{sec3} to present and resolve the classical equations of equilibrium for the spatial case. In Sect. \ref{sec4}, we compute the disturbing potential of the deformed body, the tidal force acting on the companion, and the total torque acting on the primary. The rotational equations are shown in Sect. \ref{sec5}. Then, in Sect. \ref{sec6}, we calculate the Cassini states for inviscid bodies when a uniform precession of the orbital ascending node is assumed. In Sect. \ref{sec7}, we present the conclusions. The paper is completed by three appendices, where are given some technical details of the contribution of the gravitational potential to the equilibrium equations (“Appendix 1”), a demonstration that a simple geometric composition reproduces the spatial resulting ellipsoid (“Appendix 2”) and the derivation of the $\tens{B}$-matrices (“Appendix 3”).

\section{Equilibrium equation}{\label{sec2}}

Let us consider one system formed by a rotating inviscid fluid \tens{m} of mass $m$ (primary) and a mass point \tens{M} of mass $M$ (companion), and let $\vec{r}$ the radius vector of \tens{M} in a system of reference centered on \tens{m}. Let us also consider the case that the spin rate vector of the primary, denoted by $\vec{\omega}$, is non-perpendicular to the orbital plane of the companion. We also assume that the extended primary is a homogeneous body.

\begin{figure}
\begin{center}
\includegraphics[scale=0.5]{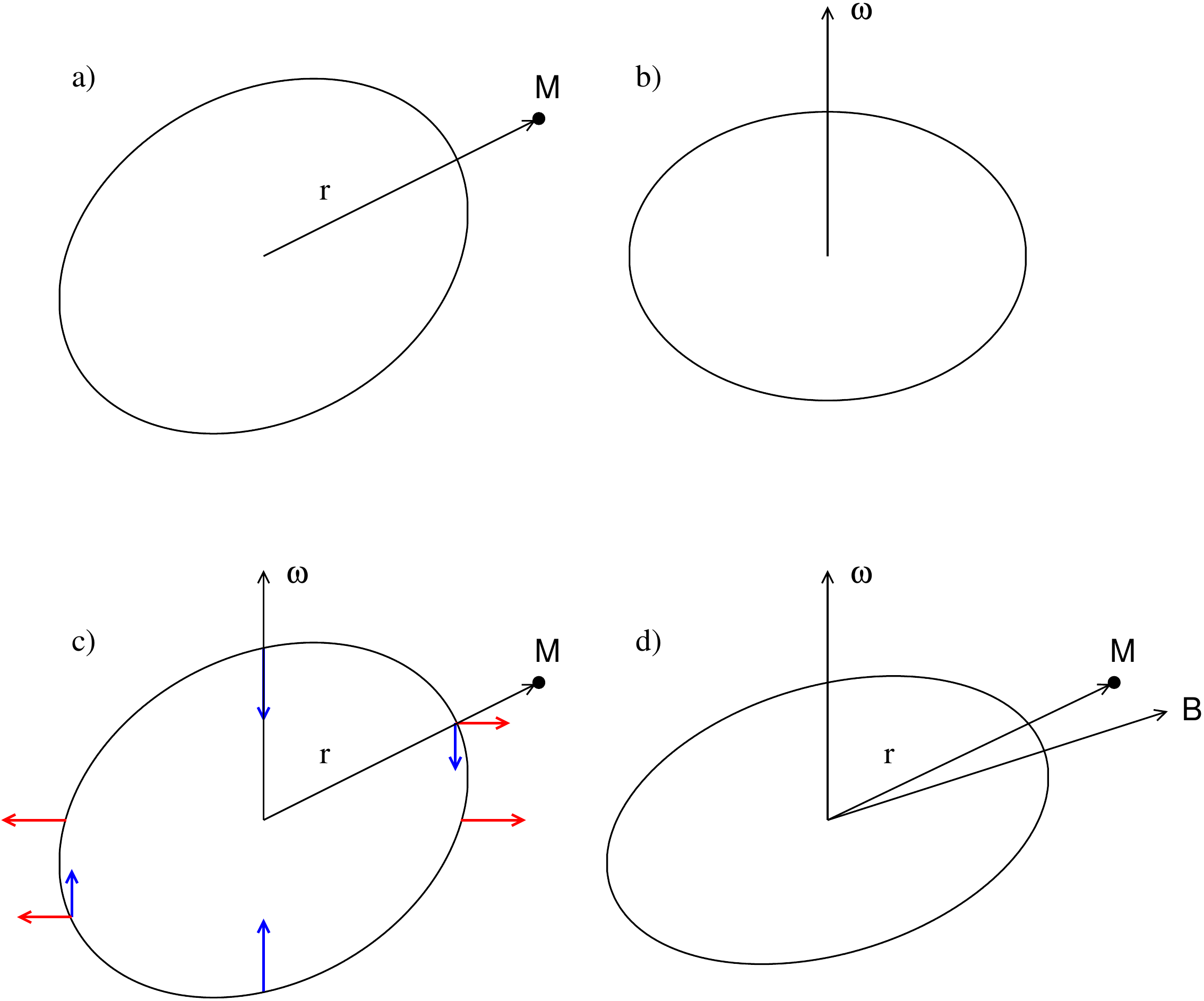}
\caption{\textit{a)} A Jeans prolate spheroid. The equatorial vertex point toward the companion. \textit{b)} A Maclaurin oblate spheroid. \textit{c)} The rotation deforms a Jeans ellipsoid, for example, contracts each point of the surface of the primary in the spin direction and distends in the perpendicular directions (blue and red arrows, respectively). \textit{d)} The equilibrium ellipsoid resulting when both the rotation and the tide are considered. The equatorial vertex does not point toward the companion if the radius vector $\vec{r}$ is not perpendicular to the spin vector $\vec{\omega}$.
}
\label{fig01}
\end{center}
\end{figure}

In the absence of rotation, the primary acquires the form of a prolate ellipsoid, whose vertex points toward the companion (panel \textit{a} of Fig. \ref{fig01}). On the other hand, in the absence of the companion, the rotation deforms the primary, flattening the body in the spin direction and distending at the equator (panel \textit{b} of Fig. \ref{fig01}). In both cases, to the first order in the deformation, the resulting figures are ellipsoids of revolution whose symmetry axes are the radius-vector or the spin rate vector, respectively. They are often called Jeans and Maclaurin ellipsoids (Chandrasekhar, 1969). 

If we start considering the spin rate acting on a Jeans ellipsoid, the rotation contracts each point of the surface of the primary in the spin direction and distends in the perpendicular directions (blue and red arrows, respectively, in panel \textit{c} of Fig. \ref{fig01}). The resulting figures are a triaxial ellipsoid with semiaxes $a_{\rm m}>b_{\rm m}>c_{\rm m}$, where the vertex does not necessarily point toward the companion, but the semiaxes $a_{\rm m}$ and $c_{\rm m}$ necessarily must lie in the plane defined by the vectors $\vec{r}$ and $\vec{\omega}$ (panel \textit{d} of Fig. \ref{fig01}). In the spatial case, both the equilibrium shape and its orientation are unknown and must be calculated.

\begin{figure}
\begin{center}
\includegraphics[scale=0.5]{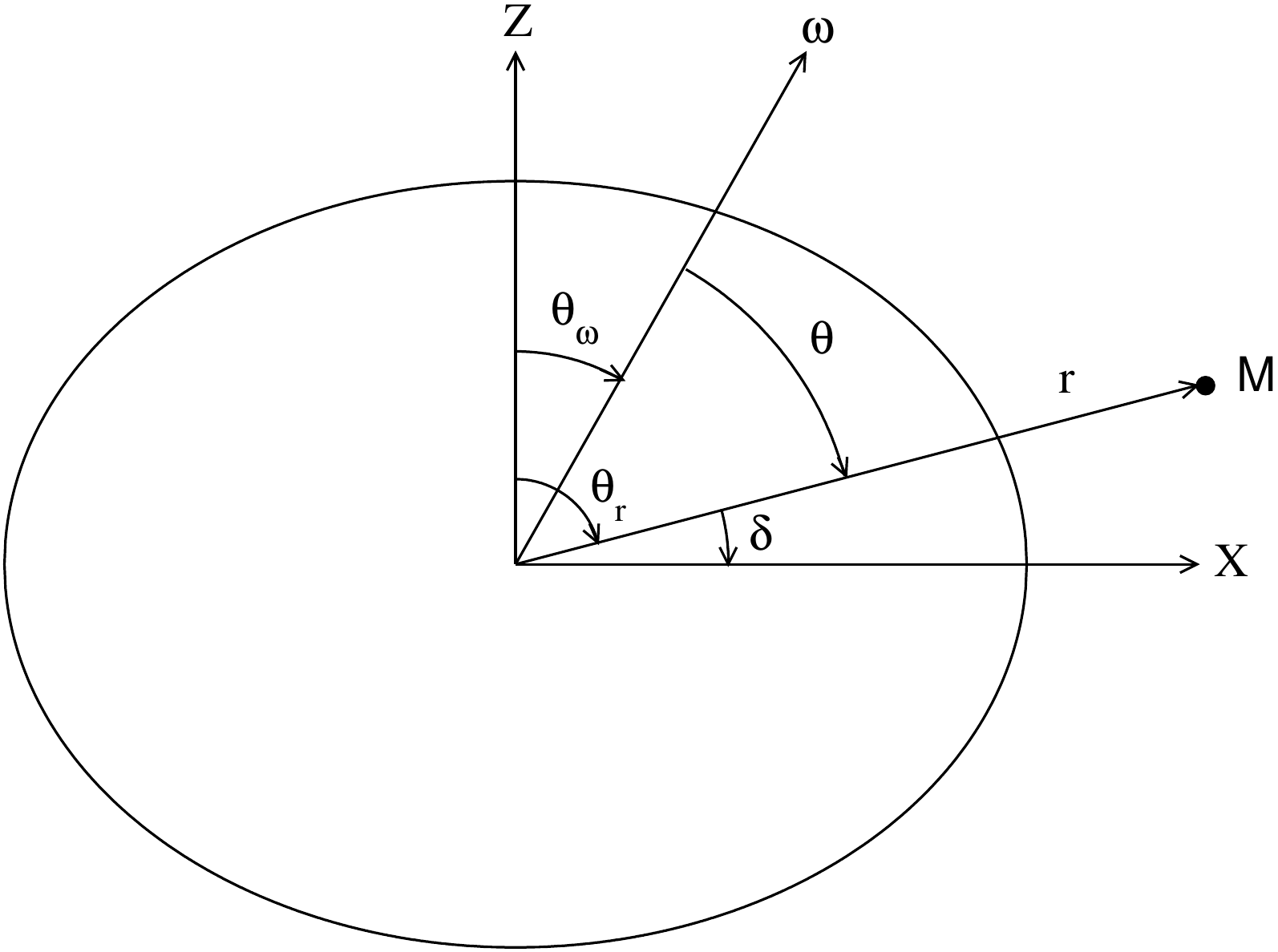}
\caption{The reference system $\mathcal{F}_\mathcal{B}$ fixed to the principal axes of inertia. The angles $\theta_\omega$ and $\theta_r$ are the co-latitudes in $\mathcal{F}_\mathcal{B}$ of the spin vector $\vec{\omega}$ and the radius vector $\vec{r}$, respectively. $\theta$ is the angle between the vectors $\vec{\omega}$ and $\vec{r}$, and the orientation angle $\delta$ is the angle between $\vec{r}$ and $\widehat{X}$.}
\label{fig02}
\end{center}
\end{figure}

In order to proceed, we consider a reference system $\mathcal{F}_\mathcal{B}$ fixed to the principal axes of inertia, with origin at the center of \tens{m} and unitary vectors $(\widehat{X},\widehat{Y},\widehat{Z})$, where $\widehat{Y}=\vec{\omega}\times\vec{r}/|\vec{\omega}\times\vec{r}|$ and $\widehat{X}$ points toward the equatorial vertex such that $(\widehat{X}\cdot \vec{r}) \ge 0$. In this reference system, the spin rate vector and the radius-vector, respectively, can be written as
\begin{equation}
\vec{\omega}=\omega\sin{\theta_\omega}\widehat{X}+\omega\cos{\theta_\omega}\widehat{Z};\qquad\qquad 
\vec{r}=r\sin{\theta_r}\widehat{X}+r\cos{\theta_r}\widehat{Z},
\label{eq:r}
\end{equation}
where $\theta_\omega$ and $\theta_r$ are the angles between $\widehat{Z}$ and the vectors $\vec{\omega}$ and $\vec{r}$, respectively, $\omega=|\vec{\omega}|$ and $r=|\vec{r}|$ (see Fig. \ref{fig02}). We also define the angles $\theta$, $\theta_\mathcal{B}$ and $\delta$ as
\begin{equation}
\theta = \theta_r-\theta_\omega;\qquad\qquad 
\delta = \frac{\pi}{2}-\theta_r;\qquad\qquad  \theta_\mathcal{B}=\frac{\pi}{2}-\theta_\omega=\theta+\delta.
\label{eq:theta-ome}
\end{equation}

Considering one point on the surface of the body $\vec{X}=X\widehat{X}+Y\widehat{Y}+Z\widehat{Z}$, the equilibrium equation can be written as 
\begin{equation}
\nabla \Phi\propto \nabla V_G + \vec{\omega}\times(\vec{\omega}\times\vec{X}),
\label{eq:equi}
\end{equation}
(e.g., Tisserand, 1891; Jardetzky, 1958; Chandrasekhar, 1969; Folonier et al. 2015), where
\begin{equation}
 \Phi = \frac{X^2}{a_{\rm m}^2}+\frac{Y^2}{b_{\rm m}^2}+\frac{Z^2}{c_{\rm m}^2}-1=0,
\label{sup}
\end{equation}
is the equation of the surface of the ellipsoid, $V_G$ is the potential of the gravitational forces at $\vec{X}$ and the second term on the right hand of (\ref{eq:equi}) corresponds to the centrifugal force.

The equilibrium equation, given by Eq. (\ref{eq:equi}), expresses the fact that the total force acting on a point of its surface must be perpendicular to the surface and can be written as:
\begin{eqnarray}
\alpha\frac{1}{Z}\Big[\vec{\omega}\times(\vec{\omega}\times\vec{X})\Big]_Z-\frac{1}{X}\Big[\vec{\omega}\times(\vec{\omega}\times\vec{X})\Big]_X &=&  \frac{1}{X}\frac{\partial V_G}{\partial X} - \alpha\frac{1}{Z} \frac{\partial V_G}{\partial Z}  \nonumber\\
\beta\frac{1}{Z}\Big[\vec{\omega}\times(\vec{\omega}\times\vec{X})\Big]_Z-\frac{1}{Y}\Big[\vec{\omega}\times(\vec{\omega}\times\vec{X})\Big]_Y&=&  \frac{1}{Y}\frac{\partial V_G}{\partial Y} - \beta\frac{1}{Z} \frac{\partial V_G}{\partial Z},
\label{eq:equilibrio1}
\end{eqnarray}
where $[\vec{a}]_{X_i}=\vec{a}\cdot\widehat{X}_i$ denotes the $X_i$ component of the vector $\vec{a}$ and
\begin{equation}
\alpha=\frac{c_{\rm m}^2}{a_{\rm m}^2}; \qquad \beta=\frac{c_{\rm m}^2}{b_{\rm m}^2}.
\end{equation}
It is important to note that the operators
\begin{equation}
\chi_X=\frac{1}{X}\frac{\partial }{\partial X} - \alpha\frac{1}{Z} \frac{\partial }{\partial Z};\qquad \chi_Y=\frac{1}{Y}\frac{\partial }{\partial Y} - \beta\frac{1}{Z} \frac{\partial }{\partial Z},
\end{equation}
are linear, so that it is possible to calculate the contribution of each gravitational potential separately.

Finally, with the constraint of volume conservation, the problem of finding the equilibrium figure (i.e., the values of the semiaxes $a_{\rm m}$, $b_{\rm m}$ and $c_{\rm m}$) is equivalent to finding the equatorial flattening $\epsilon_\rho$ and the polar oblateness $\epsilon_z$, defined as:
\begin{eqnarray}
 \epsilon_\rho =\frac{a_{\rm m}-b_{\rm m}}{\sqrt{a_{\rm m}b_{\rm m}}};\qquad  \epsilon_z = 1-\frac{c_{\rm m}}{\sqrt{a_{\rm m}b_{\rm m}}}
 \label{eq:eprho-epz}.
\end{eqnarray}
Moreover, the vertex orientation is given by angle $\delta$.

\section{Flattenings and vertex orientation}{\label{sec3}}

Using the contribution of the gravitational potentials (see Eqs. \ref{eq:vm} and \ref{eq:vtid} in Appendix 1) into the equilibrium equations (\ref{eq:equilibrio1}), and neglecting terms of order 2 in the flattenings, we obtain:
\begin{eqnarray}
\omega^2\left[\cos{2\theta_\omega}+\left(\frac{X}{Z}-\frac{Z}{X}\right)\frac{\sin{2\theta_\omega}}{2}\right] &=&  \frac{4Gm}{5R^3} \left(\frac{\epsilon_\rho}{2}+\epsilon_z\right)\nonumber\\
 &&+ \frac{4Gm}{5R^3}\frac{15MR^3}{4mr^3} \left[\cos{2\theta_r} +\left(\frac{X}{Z}-\frac{Z}{X}\right)\frac{\sin{2\theta_r}}{2}\right]\nonumber\\
 \omega^2\left[\frac{1+\cos{2\theta_\omega}}{2}+\frac{X}{Z}\frac{\sin{2\theta_\omega}}{2}\right]  &=&  \frac{4Gm}{5R^3} \left(-\frac{\epsilon_\rho}{2}+\epsilon_z\right)\nonumber\\
 &&  + \frac{4Gm}{5R^3}\frac{15MR^3}{4mr^3} \left[\frac{1+\cos{2\theta_r}}{2} +\frac{X}{Z}\frac{\sin{2\theta_r}}{2}\right].
\label{eq:equilibrio2}
\end{eqnarray}

Defining $\epsilon_J$ and $\epsilon_M$ as the flattenings of the Jeans and  Maclaurin ellipsoids:
\begin{equation}
\epsilon_J = \frac{15MR^3}{4mr^3}; \qquad \epsilon_M = \frac{5R^3\omega^2}{4Gm},
\label{def:em}
\end{equation}
and using Eq. (\ref{eq:theta-ome}), the equatorial flattening $\epsilon_\rho$ and the polar oblateness $\epsilon_z$ of the primary can be written as
\begin{eqnarray}
\epsilon_\rho &=& \frac{\epsilon_J-\epsilon_M}{2}+\frac{1}{2}\Big(\epsilon_J\cos{2\delta}-\epsilon_M\cos{2\theta_\mathcal{B}}\Big)+\frac{1}{2}\frac{Z}{X}\Big(\epsilon_J\sin{2\delta}-\epsilon_M\sin{2\theta_\mathcal{B}}\Big)\nonumber\\
\epsilon_z   &=& \frac{\epsilon_M-\epsilon_J}{4}+\frac{3}{4}\Big(\epsilon_J\cos{2\delta}-\epsilon_M\cos{2\theta_\mathcal{B}}\Big)-\frac{1}{4}\left(\frac{2X}{Z}-\frac{Z}{X}\right)\Big(\epsilon_J\sin{2\delta}-\epsilon_M\sin{2\theta_\mathcal{B}}\Big),
\label{eq:achata2}
\end{eqnarray}
where $\theta_\mathcal{B}=\theta+\delta$. 

The flattenings cannot depend on the coordinates $X$ or $Z$, and hence Eqs. (\ref{eq:achata2}) can be written as
\begin{eqnarray}
\epsilon_\rho &=& \frac{\epsilon_J}{2}-\frac{\epsilon_M}{2}+\frac{1}{2}\sqrt{\epsilon_J^2+\epsilon_M^2
 -2\epsilon_J\epsilon_M\cos{2\theta}}\nonumber\\
\epsilon_z    &=& \frac{\epsilon_M}{4}-\frac{\epsilon_J}{4}+\frac{3}{4}\sqrt{\epsilon_J^2+\epsilon_M^2
 -2\epsilon_J\epsilon_M\cos{2\theta}},
\label{eq:achata3}
\end{eqnarray}
with
\begin{eqnarray}
 \epsilon_J\cos{2\delta}-\epsilon_M\cos{2\theta_\mathcal{B}} &=& \sqrt{\epsilon_J^2+\epsilon_M^2
 -2\epsilon_J\epsilon_M\cos{2\theta}}\nonumber\\
 \epsilon_J\sin{2\delta}-\epsilon_M\sin{2\theta_\mathcal{B}} &=& 0.
\label{eq:momentum}
\end{eqnarray}

Finally, imposing the planar conditions $\epsilon_\rho(\theta=\pi/2) = \epsilon_J$ and $ \epsilon_z(\theta=\pi/2) = \frac{\epsilon_J}{2}+\epsilon_M$, the orientation angle $\delta$ can be written in terms of $\theta$ as:
\begin{eqnarray}
\cos{2\delta}&=&\frac{\epsilon_J-\epsilon_M\cos{2\theta}}{\sqrt{\epsilon_J^2+\epsilon_M^2
 -2\epsilon_J\epsilon_M\cos{2\theta}}}\nonumber\\
 \sin{2\delta}&=&\frac{\epsilon_M\sin{2\theta}}{\sqrt{\epsilon_J^2+\epsilon_M^2
 -2\epsilon_J\epsilon_M\cos{2\theta}}}.
\label{eq:delta}
\end{eqnarray}
\begin{figure}[h]
	\begin{center}
		\includegraphics[scale=0.35]{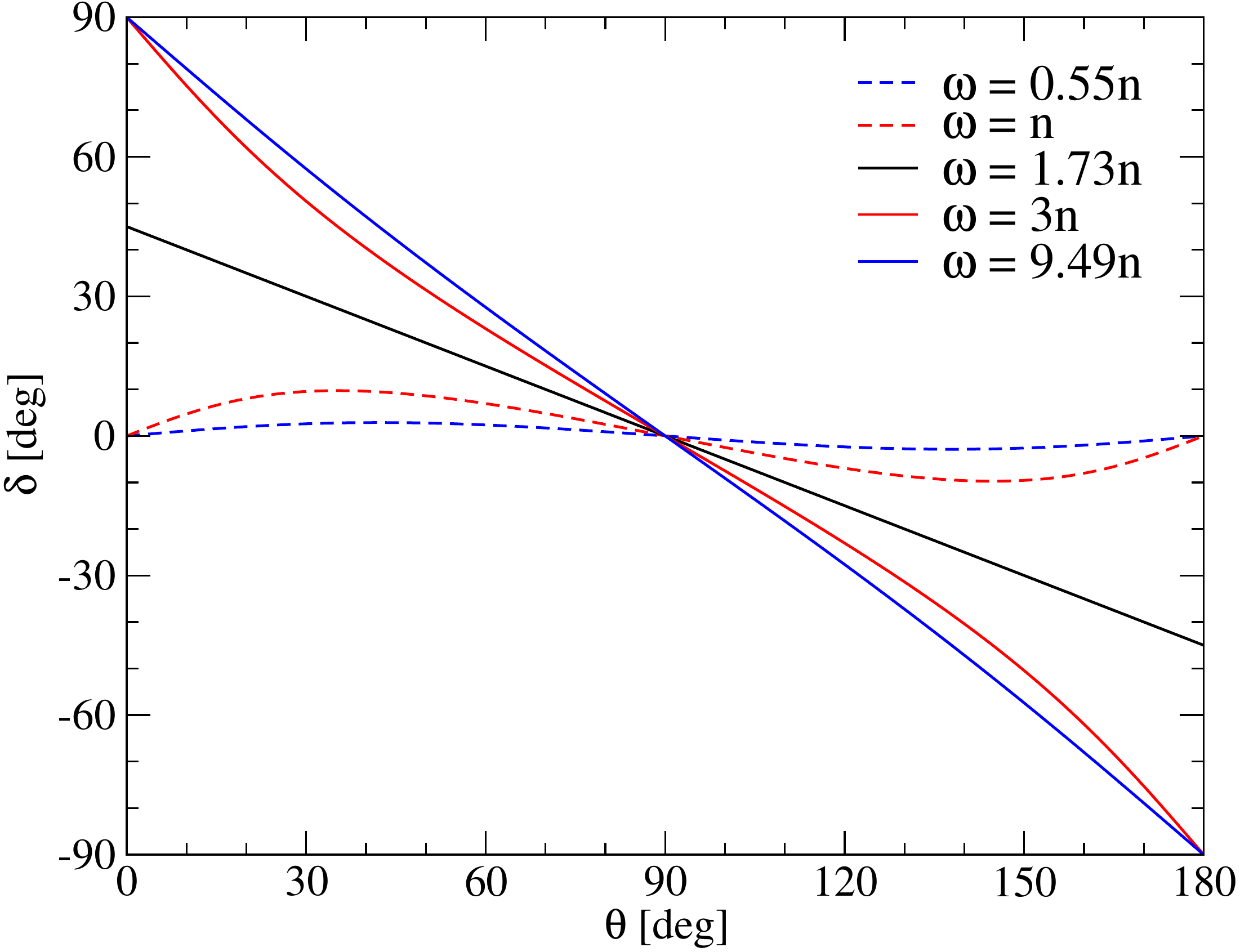}
		\caption{The orientation angle $\delta$ as a function of $\theta$ for different values of $\omega = \sqrt{3\kappa}n$ ($\kappa=1/10$ dashed blue line, $\kappa=1/3$ dashed red line, $\kappa=1$ solid black line, $\kappa=3$ solid red line and $\kappa=10$ solid blue line).}
		\label{fig03}
	\end{center}
\end{figure}

Figure \ref{fig03} shows the orientation of the vertex $\delta$ as a function of the angle $\theta$, for different values of $\kappa$, defined as 
\begin{equation}
\kappa=\frac{\epsilon_M}{\epsilon_J}=\frac{r^3\omega^2}{3GM}\approx\frac{\omega^2}{3n^2}.
\label{eq:kappa}
\end{equation}

In one orbital period, the angles $\theta$ and $\delta$ oscillate around $\pi/2$ and 0, respectively, with amplitudes $A_\theta$ and $A_\delta$, respectively. The amplitude of oscillation of the orientation angle $\delta$ depends on $A_\theta$ and $\kappa$. 

If $\kappa\geq1$, or the spin rate is $\omega\geq\sqrt{3}n$ (solid lines in Figure \ref{fig03}), $A_\delta$ increases if $A_\theta$ or $\kappa$ increases, but always satisfying $A_\delta<A_\theta$. In the limit $\kappa\rightarrow\infty$, we obtain the $A_\delta=A_\theta$. In this case, the tide (or $\epsilon_J\ll\epsilon_M$) is negligible compared to the rotation.

\begin{figure}[h]
	\begin{center}
		\includegraphics[scale=0.34]{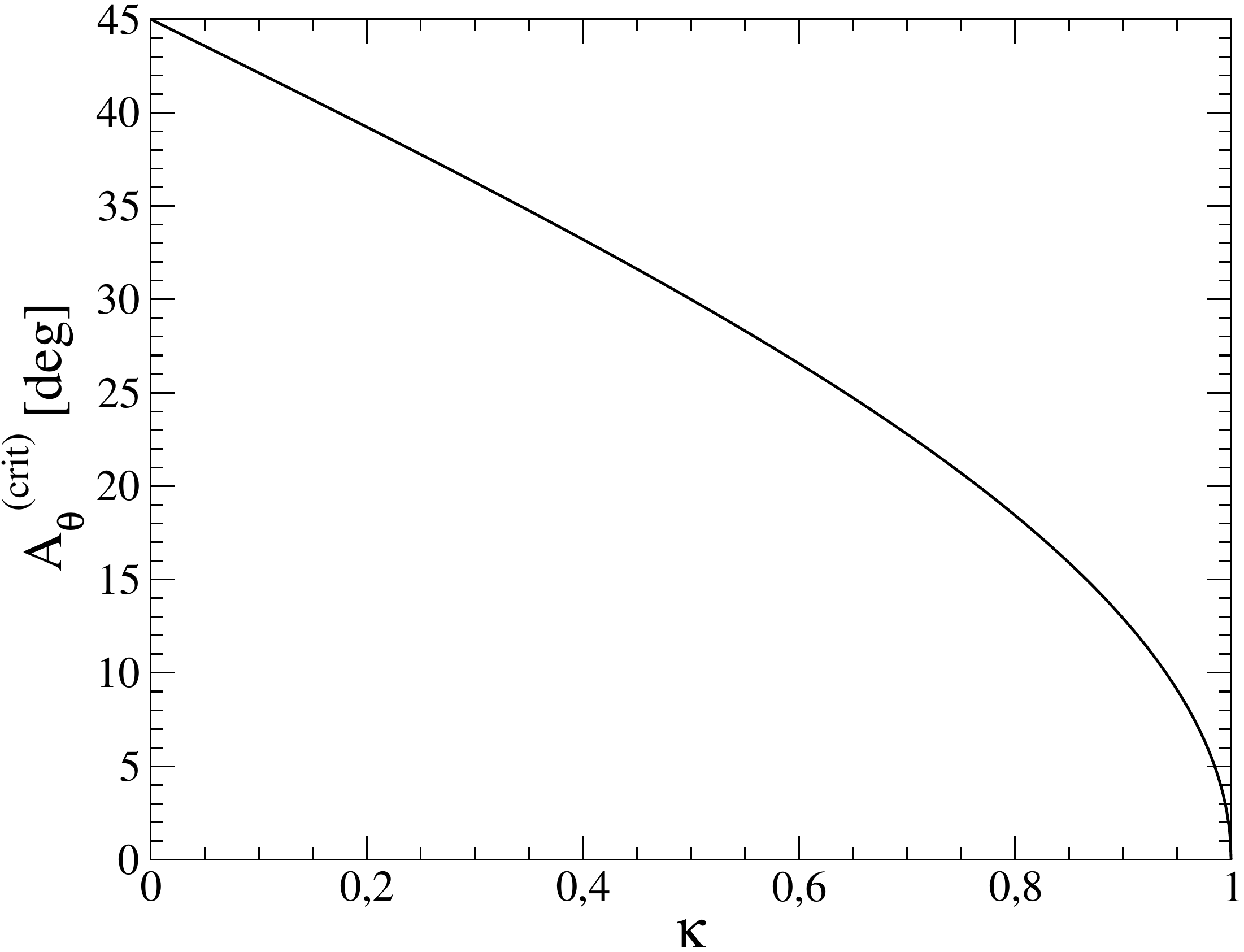}
		\includegraphics[scale=0.34]{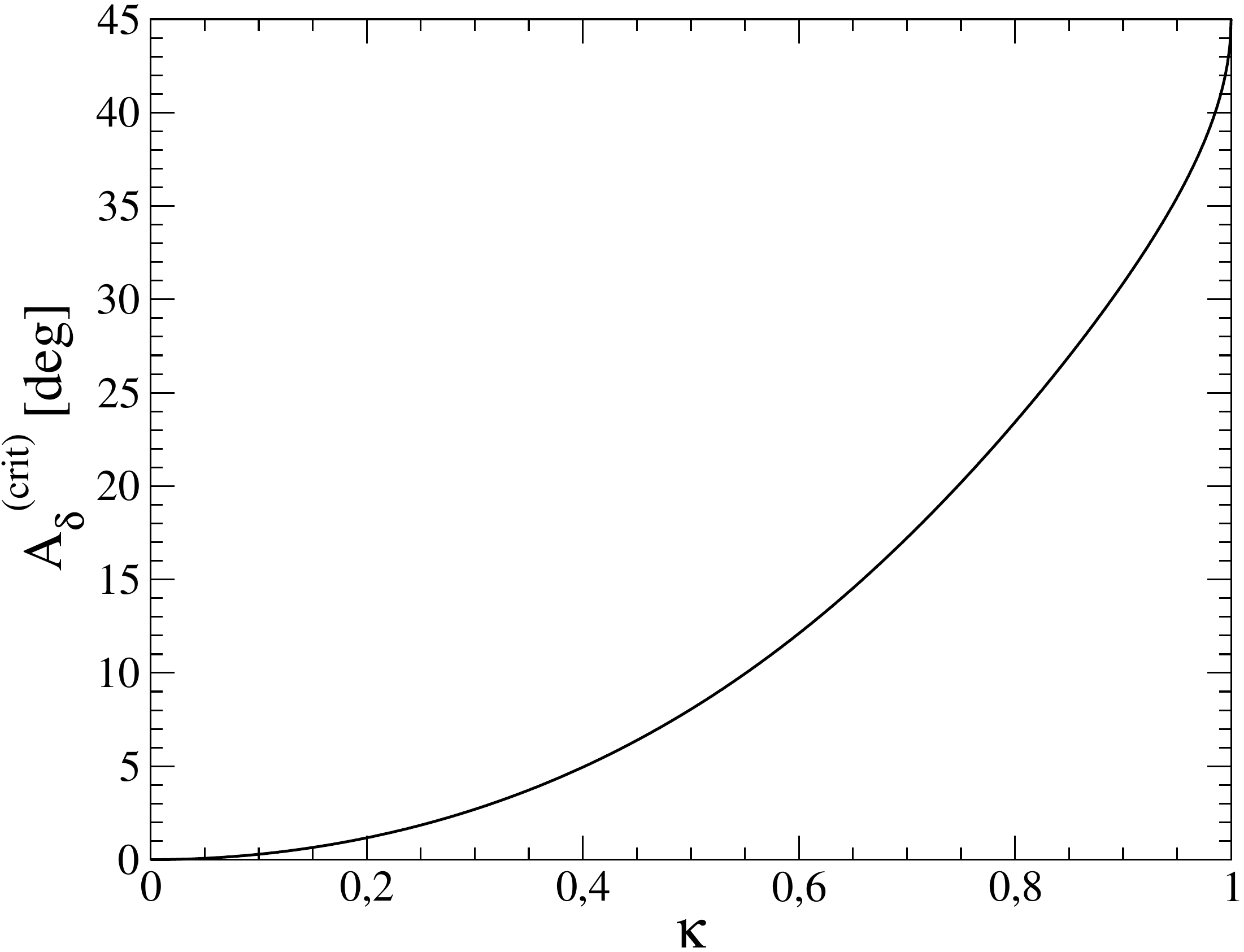}
		\caption{The critical angles $A_\theta^{(crit)}$ (\textit{left}) and $A_\delta^{(crit)}$ (\textit{right}) as a function of $\kappa$.}
		\label{fig04}
	\end{center}
\end{figure}

If $\kappa<1$ (dashed lines in Figure \ref{fig03}), $A_\delta$ increases if $A_\theta$ or $\kappa$ increases, until a critical value 
$$A_\theta^{(crit)}=\frac{1}{2}\sin^{-1}{(1-\kappa^2)},$$
from which $A_{\delta}$ decreases if $A_{\theta}$ increases. The critical value $A_\theta^{(crit)}$ and its corresponding $A_\delta^{(crit)}$ depend on $\kappa$
$$A_\delta^{(crit)}=\frac{1}{2}\tan^{-1}{\left(\frac{\kappa^2}{\sqrt{1-\kappa^2}}\right)}.$$
In the limit $\kappa\rightarrow0$, we obtain $A_\delta=0$. In this case, the rotation is negligible compared to the tide (or $\epsilon_J\gg\epsilon_M$) and the equatorial vertex points toward the companion. Critical amplitudes $A_\theta^{(crit)}$ and $A_\delta^{(crit)}$ are shown in Figure \ref{fig04}.

\begin{figure}[h]
	\begin{center}
		\includegraphics[scale=0.34]{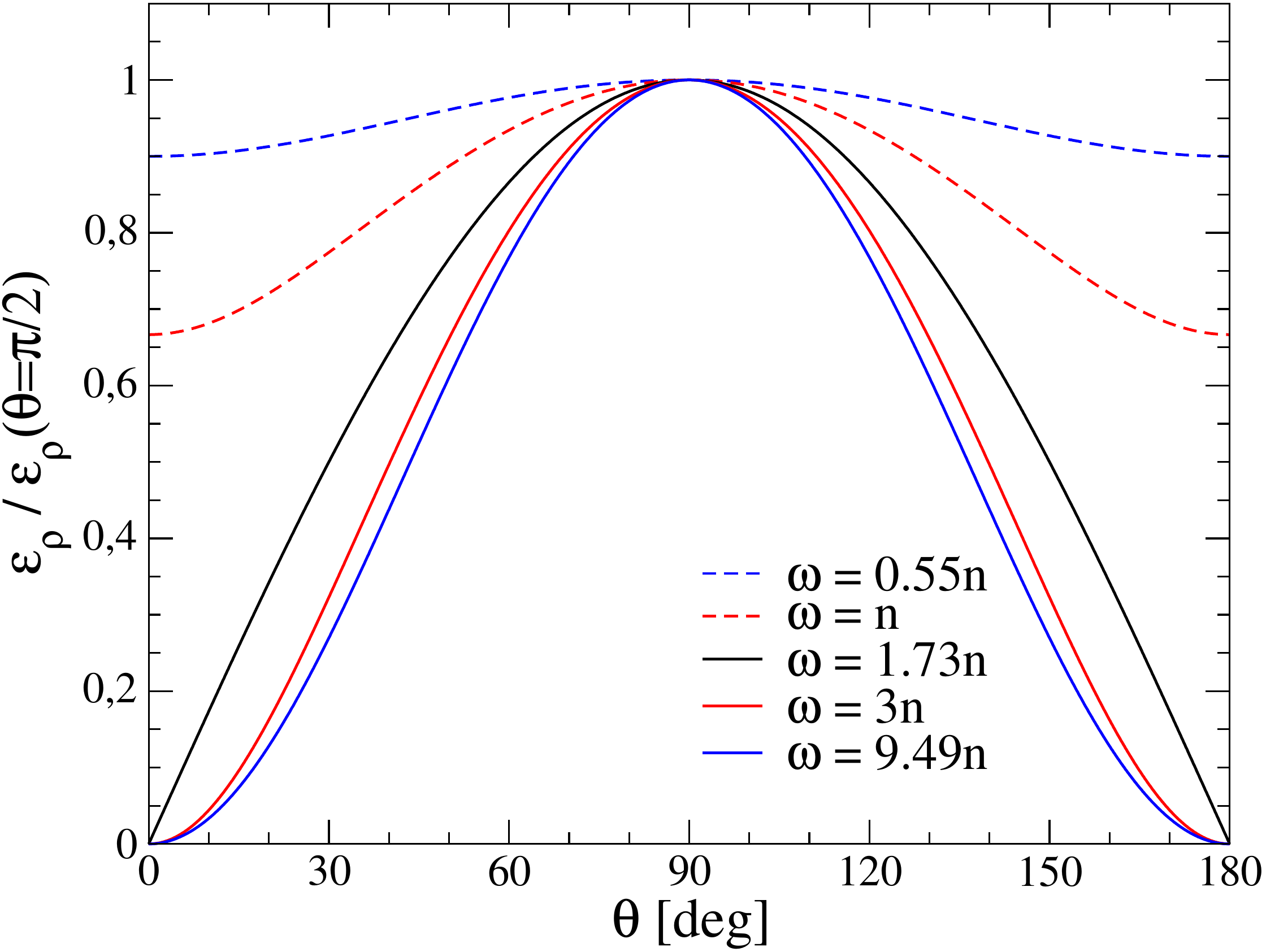}
		\includegraphics[scale=0.34]{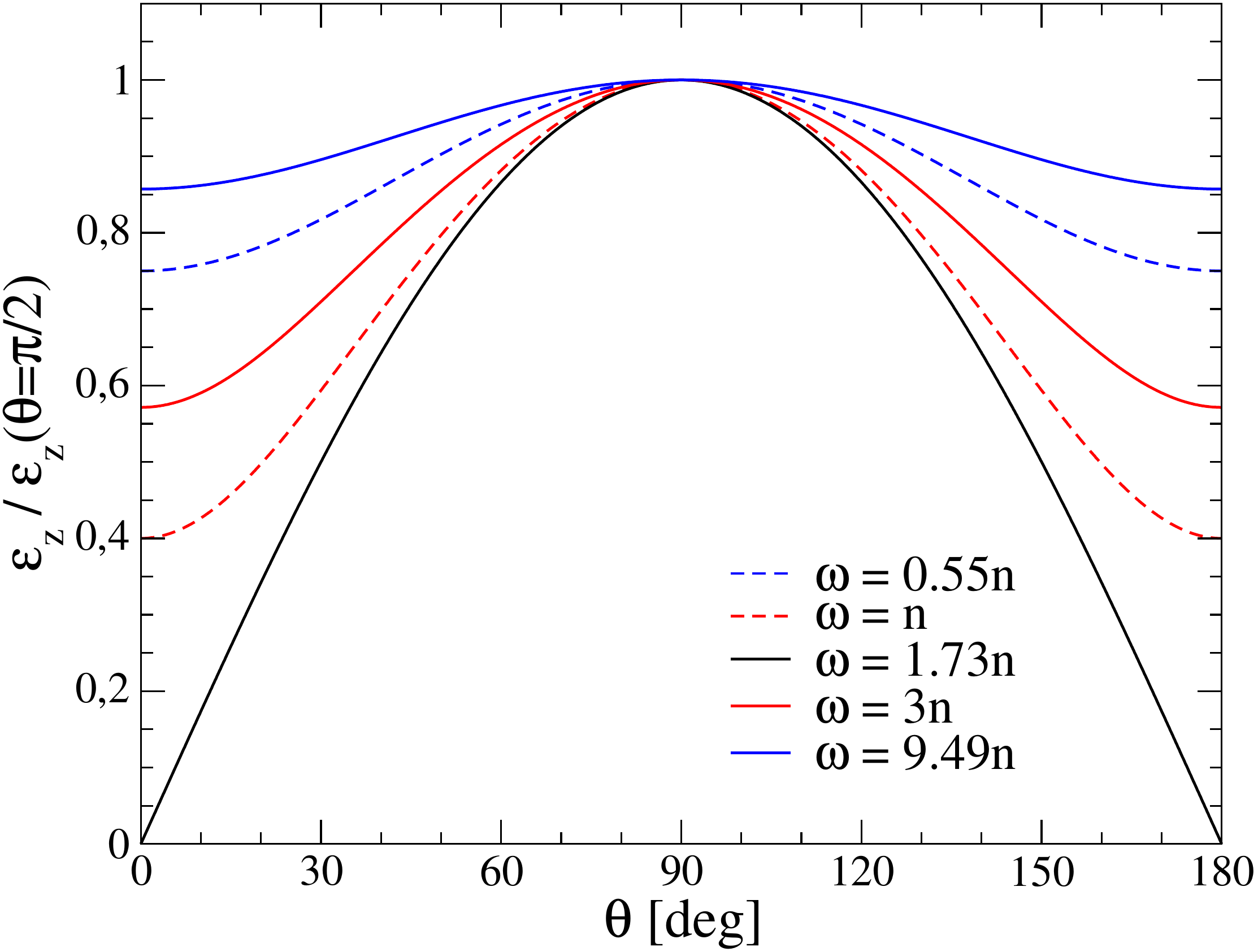}
		\caption{The normalized equatorial flattening $\epsilon_\rho$ (\textit{left}) and the normalized polar flattening $\epsilon_z$ (\textit{right}) as a function of $\theta$ for different values of $\omega = \sqrt{3\kappa}n$ ($\kappa=1/10$ dashed blue line, $\kappa=1/3$ dashed red line, $\kappa=1$ solid black line, $\kappa=3$ solid red line and $\kappa=10$ solid blue line).}
		\label{fig05}
	\end{center}
\end{figure}
In Figure \ref{fig05} we show the equatorial flattening (\textit{left} panel) and the polar flattening (\textit{right} panel) are a function of $\theta$, each one of these flattenings normalized to its respective value in the planar case $\theta=\pi/2$. 

\begin{figure}[h]
	\begin{center}
		\includegraphics[scale=0.35]{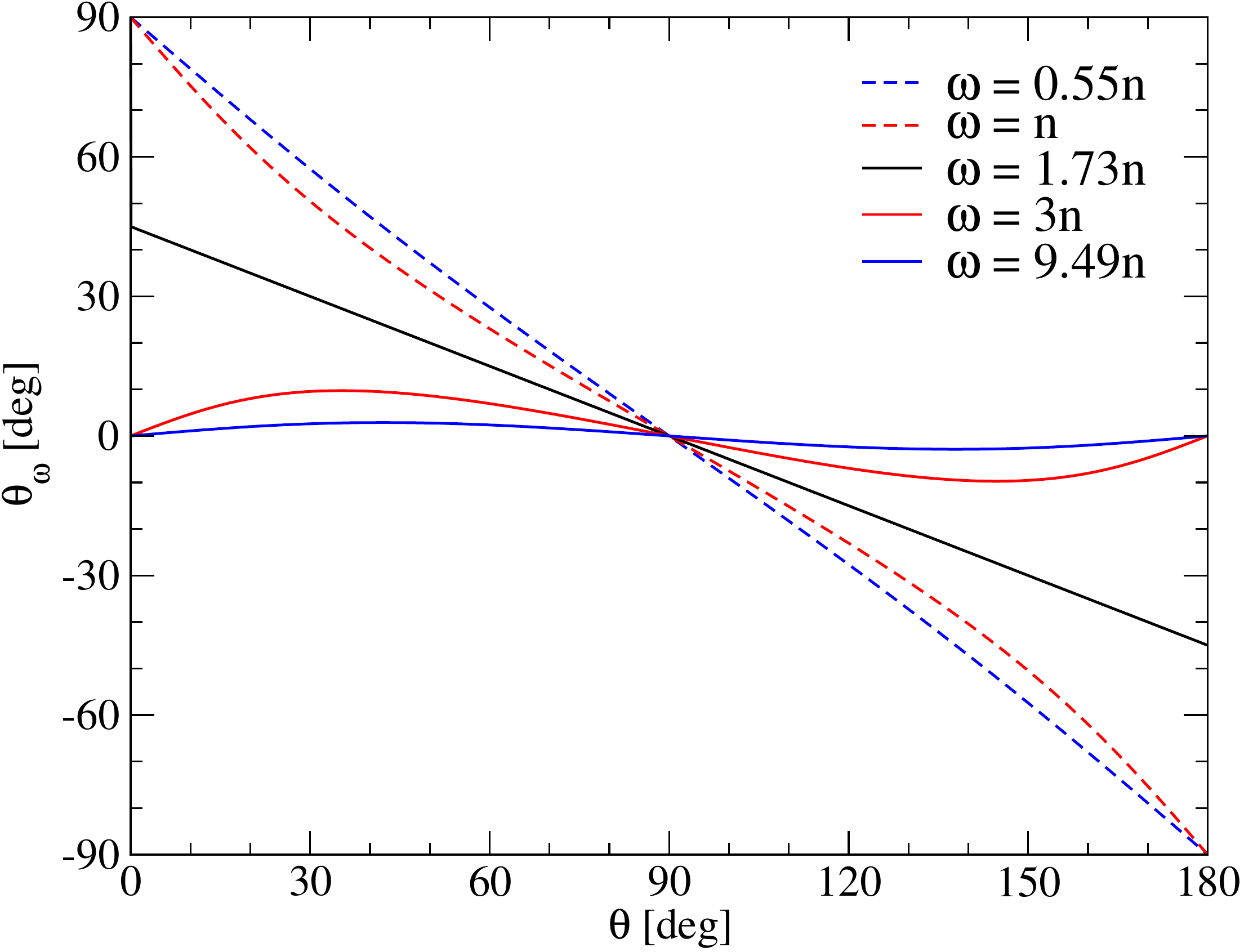}
		\caption{The angle $\theta_\omega$ as a function of $\theta$ for different values of $\omega = \sqrt{3\kappa}n$ ($\kappa=1/10$ dashed blue line, $\kappa=1/3$ dashed red line, $\kappa=1$ solid black line, $\kappa=3$ solid red line and $\kappa=10$ solid blue line).}
		\label{fig06}
	\end{center}
\end{figure}

Finally, in Figure \ref{fig06}, the angle
$$\theta_\omega=\frac{\pi}{2}-\theta-\delta,$$ 
is shown as a function of $\theta$ for different values of $\kappa$. This angle measures the angular distance between the direction of the principal axis of inertia $C$ and the direction of the spin axis. In the planar case, where $\theta=\pi/2$, we obtain that $\theta_\omega$ is equal to zero, so the angular velocity vector is parallel to the direction of the axis of inertia. However, as $\theta$ moves away from this value, the value of $\theta_\omega$ changes. In particular, if we consider that the deformed body rotation is synchronous with the mean motion of the other body, that is $\omega=n$ (or $\kappa=1/3$), then $|\theta_\omega|$ increases when $|\theta-\pi/2|$ increases, in contrast to $\delta$, whose value is bounded by a small value ($<4^\circ$). This means that the approximation $\tens{I}\vec{\omega} \approx C\vec{\omega}$ (where \tens{I} is the inertia tensor) only makes sense in the neighborhood of the planar problem. This result is reasonable, even for dynamic tidal theories since energy dissipation tends to bring the vertex of the ellipsoid to values close to the equilibrium defined by the inviscid problem.

\section{Disturbing potential, force and torque}{\label{sec4}}

The disturbing potential created by the homogeneous triaxial ellipsoid \tens{m} at an external point $\vec{r}^*=(r^*,\theta^*,\varphi^*)$, neglecting the harmonics of degree higher than 2, is:
\begin{equation}
 \delta U(\vec{r}^*) = \frac{3G}{2r^{*5}}\Big(\vec{r}^*\cdot \tens{I}\vec{r}^*-\frac{r^{*2}}{3}\textnormal{Tr}(\tens{I})\Big),
\end{equation}
(see, e.g., Beutler, 2005; Murray and Dermott, 1999; Correia and Rodriguez, 2013) where $\tens{I}$ is the inertia tensor, $\textnormal{Tr}(\tens{I})$ is the trace of inertia tensor, and $I_0=\frac{2}{5}mR^2=(I_{xx}+I_{yy}+I_{zz})/3=\textnormal{Tr}(\tens{I})/3$.

As we show in appendices 2 and 3, if the deformations are small enough, we can linearize the inertia tensor in terms of equatorial flattening and polar flattening, so it is possible to separate tidal and rotational components so that:
\begin{equation}
 \tens{I} = I_0(\mathbb{I}-\tens{B}_M-\tens{B}_J).
\end{equation}
where $\mathbb{I}$ is the identity matrix and $\tens{B}$-matrices
\begin{equation}
 \tens{B}_M = \frac{\epsilon_M}{\omega^2}\Big(-\vec{\omega}\otimes\vec{\omega}+\frac{\omega^2}{3}\mathbb{I}\Big); \qquad \tens{B}_J =\frac{\epsilon_J}{r^2}\Big(\vec{r}\otimes\vec{r}-\frac{r^2}{3}\mathbb{I}\Big),
 \label{eq:BM-BJ}
\end{equation}
are the non-dimensional form of the moment of quadrupole matrix due to the rotational and the tidal deformation (Ragazzo and Ruiz, 2015; Ragazzo and Ruiz, 2017; Boué, 2020).

To obtain the force $\vec{F}$ acting on the companion \tens{M}, it is enough to make the identification $\vec{r}^*$ by $\vec{r}$ and $M^*$ by $M$ after the calculation of the gradient of the disturbing potential, thus:
\begin{equation}
\vec{F}= -M\nabla_{\vec{r}^*} \delta U(\vec{r}^*)\Big|_{\vec{r}^*=\vec{r}}.
\end{equation}
The sign in this expression comes from the fact that we are using the conventions of Physics ($\delta U$ is a potential, not a force function). Then, we obtain:
\begin{equation}
\vec{F}=-\frac{3GM}{2r^5}\Big(2\tens{I}\vec{r}-\frac{5\vec{r}}{r^2}\vec{r}\cdot \tens{I}\vec{r}+\textnormal{Tr}(\tens{I})\vec{r}\Big).
\label{eq:F}
\end{equation}

Finally, the corresponding torque acting on the primary is $\vec{M}=-\vec{r}\times\vec{F}$, or
\begin{equation}
\vec{M}=\frac{3GM}{r^5}\vec{r}\times\tens{I}\vec{r}.
\end{equation}

Using the \tens{B}-matrices given by Eq. (\ref{eq:BM-BJ}), the torque can be written as:
\begin{equation}
\vec{M}=-\frac{3GMI_0}{r^5}\vec{r}\times\tens{B}_M\vec{r}=\frac{3GMI_0}{r^5}\epsilon_M(\vec{r}\cdot\vec{s})(\vec{r}\times\vec{s}),
\end{equation}
where $\vec{s}=\vec{\omega}/\omega$ is the spin pole.

\section{Rotational equation}{\label{sec5}}

In order to obtain the rotational equation for the spin vector $\vec{\omega}$, we consider the angular momentum
\begin{equation}
 \vec{L} = \tens{I}\vec{\omega},
\end{equation}
and its derivative with respect to time
\begin{equation}
 \dot{\vec{L}} = I_0\frac{d}{dt}\Big[(\mathbb{I}-\tens{B}_M-\tens{B}_J)\vec{\omega}\Big].
 \label{eq:dotL}
\end{equation}

Defining the auxiliary matrix $\tens{U}_M$ as
\begin{equation}
\tens{U}_M = -\frac{4\epsilon_M}{3\omega^2}\Big(\vec{\omega}\otimes\vec{\omega}+\frac{\omega^2}{2}\mathbb{I}\Big),
\end{equation}
and using Eq. (\ref{eq:BM-BJ}), the derivative with respect to time of the term $\tens{B}_M\vec{\omega}$ can be written as
\begin{equation}
\frac{d}{dt}\big(\tens{B}_M\vec{\omega}\big)=\tens{U}_M\dot{\vec{\omega}}.
\end{equation}

Neglecting second-order contributions in flattening
\begin{eqnarray}
(\mathbb{I}-\tens{U}_M-\tens{B}_J)^{-1}&=&(\mathbb{I}+\tens{U}_M+\tens{B}_J)+\mathcal{O}(\epsilon^2)\nonumber\\
(\tens{U}_M+\tens{B}_J)\dot{\tens{B}}_J &=& \mathcal{O}(\epsilon^2)\nonumber\\
(\tens{U}_M+\tens{B}_J)\vec{M} &=& \mathcal{O}(\epsilon^2),
\label{eq:approx}
\end{eqnarray}
finally, the differential equation for the angular velocity vector, resulting from Eqs. (\ref{eq:dotL})-(\ref{eq:approx}), at first order in flattening and in an inertial frame, is:
\begin{equation}
 \vec{\dot{\omega}} = \frac{3GM}{r^5}\epsilon_M(\vec{r}\cdot\vec{s})(\vec{r}\times\vec{s}) + \dot{\tens{B}}_J\vec{\omega},
\label{eq:dot-omega}
\end{equation}
where the derivative $\dot{\tens{B}}_J$ can be written in terms of the radius vector $\vec{r}$ of \tens{M} and its velocity $\vec{V}$
\begin{equation}
\dot{\tens{B}}_J = \frac{\epsilon_J}{r^2}\Big(\vec{V}\otimes\vec{r}+\vec{r}\otimes\vec{V}-\frac{5\vec{r}\cdot\vec{V}}{r^2}\vec{r}\otimes\vec{r}+\vec{r}\cdot\vec{V}\ \mathbb{I}\Big).
\end{equation}

\section{The Cassini state of an inviscid body}{\label{sec6}}

In this Section, we are interested in the Cassini equilibrium configurations for the case of an inviscid body. These are the fixed points of the spin axis in the frame rotating at the precession frequency of the orbital plane. Following the theory presented by Colombo (1966), we remove the short-period of the differential equation for the spin rate, given by Eq. (\ref{eq:dot-omega}), taking the time-average over one period assuming the spin rate vector $\vec{\omega}$ constant. For this calculation, we consider the two-body Keplerian motion:
\begin{equation}
 r=\frac{a(1-e^2)}{1+e\cos{v}}.
\end{equation}
Here $a,e,v$ are the semi-major axis, the eccentricity and the true anomaly, respectively.

In order to perform the averaging, we also consider a reference frame where the $x-y$ plane is the orbital plane and the $x$-axis along the direction of the periapsis. Thus, the radius-vector $\vec{r}$ and the velocity vector $\vec{V}$ of the companion are:
\begin{equation}
 \vec{r}=r\left(\begin{array}{c}
 \cos{v} \\
 \sin{v} \\
 0  \end{array} \right);\qquad
 \vec{V}=\frac{na}{\sqrt{1-e^2}}\left(\begin{array}{c}
-\sin{v} \\
 e+\cos{v} \\
 0  \end{array} \right).
 \label{eq:r-v}
\end{equation}

Hence, using the radius-vector $\vec{r}$ and the velocity $\vec{V}$ given by Eq. (\ref{eq:r-v}), the time-average over one period of each term of the differential Eq. (\ref{eq:dot-omega}) for the spin rate is:\footnote{The average over one orbital period of one function $f$ can be written as:
$$
\langle f\rangle = \frac{1}{P}\int_0^Pf(v(t))\ dt = \frac{1}{2\pi}\int_0^{2\pi}f(v)\ \frac{r^2dv}{a^2\sqrt{1-e^2}},
$$
where the integration is done over the true anomaly $v$ instead of being done over the mean anomaly $\ell$, via the classical expression $r^2\ dv = a^2\sqrt{1-e^2}\ d\ell$.}
\begin{eqnarray}
\left\langle\frac{3GM}{r^5}\epsilon_M(\vec{r}\cdot\vec{s})(\vec{r}\times\vec{s})\right\rangle &=& \frac{3GM}{a^3}\frac{\epsilon_M}{2(1-e^2)^{3/2}}(\vec{s}\cdot\vec{n})(\vec{s}\times\vec{n})\nonumber\\
\left \langle\dot{\tens{B}}_J\vec{\omega}\right\rangle &=& 0,
\end{eqnarray}
where $\vec{n}$ is the orbit pole.

Finally, we consider that the orbital plane is precessing uniformly with constant inclination $I$ at the rate $g$ around the normal $\vec{k}$ of the invariant plane (also called Laplace plane). Using the reference frame $\mathcal{F}_g=(\vec{i},\vec{j},\vec{k})$ centered on the primary and with the $z$-axis and the $x$-axis pointing to $\vec{k}$ and to the direction of the ascending node, respectively, the differential equation for the spin pole can be written as:
\begin{eqnarray}
\dot{\vec{s}}+g(\vec{k}\times\vec{s})&=& \alpha(\vec{s}\cdot\vec{n})(\vec{s}\times\vec{n}),
\label{eq:dot-s-Fg}
\end{eqnarray}
where the constant $\alpha$ is:
\begin{equation}
\alpha = \frac{3GM}{a^3}\frac{\epsilon_M}{2\omega(1-e^2)^{3/2}}.
\label{eq:alpha}
\end{equation}

Considering the contribution of the rotation to the principal moment of inertia $\epsilon_M = \big((C-A)/C\big)_{\rm rot}$, and the approximation $GM/a^3\approx n^2$, the coefficient $\alpha$ can be written as
\begin{equation}
\alpha = \frac{3n^2}{2\omega(1-e^2)^{3/2}}\left(\frac{C-A}{C}\right)_{{\rm rot}}.
\end{equation}

The rotational equation given by Eq. (\ref{eq:dot-s-Fg}) and the $\alpha$ coefficient are same as those obtained by Colombo (1966) for a rigid axisymmetric body. Therefore, an inviscid body, deformed by the tide and by the rotation, has the same secular evolution as an axisymmetric rigid body with the same polar flattening as the one produced by the rotation only on the inviscid body.
 
The physical interpretation of this result is simple. On the one hand, unlike a triaxial rigid body as studied in Peale (1969), the equatorial vertex of a body formed by a fluid without viscosity adapts instantaneously in such a way as to always point toward the companion that creates this triaxiality, so the tidal contribution to the torque is canceled at every instant of time throughout the entire orbit. On the other hand, since the total torque acting on the deformed body is perpendicular to the angular velocity vector, this means that only the direction of the rotational velocity vector $\vec{s}$ changes throughout an orbit, while the spin rate $\omega$ remains constant. In this way, the polar flattening due to rotation remains constant throughout an orbit, in the same way as for the case of a rigid body, like the one studied in Colombo (1966). Furthermore, at first order, the effect of the variation of the inertia due to tide cancels out once averaged over one orbital period.

\section{Conclusions}{\label{sec7}}

In this paper, we found the ellipsoidal equilibrium figure of an inviscid body when its spin rate vector is not perpendicular to the plane of motion of one punctual companion. We calculated the equilibrium equations for small flattenings and found that, in contrast to the planar case, the equatorial and the polar flattenings are nonlinear functions of the Jeans and the Maclaurin flattenings and are also a function of the angle $\theta$ between the spin rate vector and the radius vector. An important consequence of this approach is that the equatorial vertex of this prolate ellipsoid does not point toward the companion as in the planar case and produces a torque perpendicular to the spin vector. On the other hand, the fact that the semiaxes of the ellipsoid are not oriented toward the external body that creates the tide, nor in the direction of the angular velocity vector, is that the principal axes of inertia do not coincide with the angular velocity vector. In this work, we find that the deviation between the principal axis of inertia $C$ and the angular velocity vector can be large, even when the angle $\delta$ is small. Thus, if the amplitude of $\theta$, over an orbital period, is large, the classical approximation $\tens{I}\vec{\omega} \approx C\vec{\omega}$ only makes sense in the neighborhood of the planar problem.

We also studied the Cassini states, induced by an external precession of the line of the orbital ascending node. Neglecting the short-period term, we calculated the differential equation for the spin direction. The equation for the spin pole is identical to those found by Colombo (1969) for a rigid axisymmetric ellipsoid. As a result of this, we conclude that an inviscid body, deformed by the tide and by the rotation, has the same secular rotational evolution as an axisymmetric rigid body with the same flattening as the one produced by the sole rotation on the inviscid body. In particular, this is also true for a short-period gas giant exoplanet whose tidal deformation has about the same amplitude as the rotation deformation.

\section*{Acknowledgments}
We wish to thank the anonymous referee for their comments and suggestions that helped to improve the manuscript. This investigation is funded by FAPESP, Grants $2016/20189-9$, $2019/11276-3$ and $2016/13750-6$ ref. PLATO mission.

\section*{Appendix 1: Gravitational potentials}{\label{ap1}}

\subsection{Potential of a homogeneous ellipsoid at an internal point}
The potential generated for the homogeneous ellipsoid on its surface, in the reference system $\mathcal{F}_\mathcal{B}$ fixed to the principal axes of inertia, is (see Kellogg, 1929)
\begin{equation}
V_{\rm m} = V_0 + A_xX^2 + A_yY^2 + A_zZ^2,
\end{equation}
where $V_0$, $A_x$, $A_y$ and $A_z$ are constants. The constants $A_{x_i}$ are
\begin{eqnarray}
A_x &=& \frac{3Gm}{4R^3} \alpha \int_0^\infty \frac{dt}{(1+\alpha t)\sqrt{(1+\alpha t)(1+\beta t)(1+t)}} \nonumber \\
A_y &=& \frac{3Gm}{4R^3} \beta  \int_0^\infty \frac{dt}{(1+\beta t) \sqrt{(1+\alpha t)(1+\beta t)(1+t)}} \nonumber \\
A_z &=& \frac{3Gm}{4R^3}        \int_0^\infty \frac{dt}{(1+t)\sqrt{(1+\alpha t)(1+\beta t)(1+t)}},
\label{const}
\end{eqnarray}
where $R$ and $m$ is the mean radius and the mass of the primary, $G$ is the gravitational constant and
\begin{equation}
\alpha=\frac{c_{\rm m}^2}{a_{\rm m}^2}=1-\epsilon_\rho-2\epsilon_z; \qquad \beta=\frac{c_{\rm m}^2}{b_{\rm m}^2}=1+\epsilon_\rho-2\epsilon_z.
\end{equation}

Hence, the contribution to the equilibrium equations, neglecting terms of order 2 in the flattenings, is
\begin{eqnarray}
\chi_X[V_{\rm m}]=\frac{1}{X}\frac{\partial V_{\rm m}}{\partial X}- \frac{\alpha}{Z}\frac{\partial V_{\rm m}}{\partial Z} &=& \frac{4Gm}{5R^3} \left(\frac{\epsilon_\rho}{2}+\epsilon_z\right)\nonumber\\
\chi_Y[V_{\rm m}]=\frac{1}{Y}\frac{\partial V_{\rm m}}{\partial Y}-\frac{\beta}{Z}\frac{\partial V_{\rm m}}{\partial Z} &=& \frac{4Gm}{5R^3} \left(-\frac{\epsilon_\rho}{2}+\epsilon_z\right).
\label{eq:vm}
\end{eqnarray}

\subsection{Tidal potential}

The potential due to the tide is (Lambeck 1980)
\begin{equation}
\nabla V_{\rm tid} = \frac{GM}{r^3}\left[\vec{X}-\frac{3(\vec{X}\cdot\vec{r})\vec{r}}{r^2}\right].
\end{equation}
Using that $\vec{r}=r\sin{\theta_r}\widehat{X}+r\cos{\theta_r}\widehat{Z}$ and $\vec{X}=X\widehat{X}+Y\widehat{Y}+Z\widehat{Z}$, we obtain
\begin{equation}
\nabla V_{\rm tid} =  \frac{GM}{r^3} \left( \begin{array}{ccc} X(1-3\sin^2{\theta_r}) - 3Z\sin{\theta_r}\cos{\theta_r} \\
Y \\
- 3X\sin{\theta_r}\cos{\theta_r}+Z(1-3\cos^2{\theta_r}) \end{array} \right).
\end{equation}

Hence, the contribution to the equilibrium equations, neglecting terms of order 2 in the flattenings, is
\begin{eqnarray}
\chi_X[V_{\rm tid}]=\frac{1}{X}\frac{\partial V_{\rm tid}}{\partial X}-\frac{\alpha}{Z}\frac{\partial V_{\rm tid}}{\partial Z} &=& \frac{4Gm}{5R^3}\epsilon_J \left[\cos{2\theta_r} +\left(\frac{X}{Z}-\frac{Z}{X}\right)\frac{\sin{2\theta_r}}{2}\right]\nonumber\\
\chi_Y[V_{\rm tid}]=\frac{1}{Y}\frac{\partial V_{\rm tid}}{\partial Y}- \frac{\beta}{Z}\frac{\partial V_{\rm tid}}{\partial Z}  &=& \frac{4Gm}{5R^3}\epsilon_J \left[\frac{1+\cos{2\theta_r}}{2} +\frac{X}{Z}\frac{\sin{2\theta_r}}{2}\right],
\label{eq:vtid}
\end{eqnarray}
where
\begin{equation}
\epsilon_J = \frac{15MR^3}{4mr^3},
\end{equation}
is the equatorial flattening of a Jeans ellipsoid.

\section*{Appendix 2: The spatial resulting ellipsoid as a simple geometric composition}{\label{ap2}}

In this appendix, we demonstrate that, at the first order in flattenings, the simple geometric composition of the Jeans and the Maclaurin ellipsoids results in the same triaxial ellipsoid calculated in Sects. \ref{sec2} and \ref{sec3}. In this way, the geometric sum of the heights of Jeans and the Maclaurin,  each of these ellipsoids with a different axis of symmetry, produces a triaxial ellipsoid with the same flattenings and the same orientation of the equilibrium ellipsoid in the spatial case. This peculiar result is a consequence of the linearity of the deformations when they are small enough.

If $\delta\rho(\widehat{\theta},\widehat{\varphi})$, $\delta\rho_{\rm tid}(\widehat{\theta},\widehat{\varphi})$ and $\delta\rho_{\rm rot}(\widehat{\theta},\widehat{\varphi})$ are the distances of the surface point of the equilibrium ellipsoid, the Jeans ellipsoid and the Maclaurin ellipsoid, respectively, of coordinates $\widehat{\varphi}$ (longitude) and $\widehat{\theta}$ (co-latitude) in the body reference frame $\mathcal{F}_\mathcal{B}$ to the sphere of radius $R$, we use the expressions
\begin{equation}
 \delta\rho = \delta\rho_{\rm tid}+\delta\rho_{\rm rot},
 \label{eq:drho-def}
\end{equation}
which is valid in the first order in flattenings. In order to proceed, we need to describe $\delta\rho$, $\delta\rho_{\rm tid}$ and $\delta\rho_{\rm rot}$ in the reference frame $\mathcal{F}_\mathcal{B}$.

We start with the description of the resulting ellipsoid $\delta\rho$. The semiaxes of this ellipsoid can be written as
\begin{equation}
 a_{\rm m} = R\Big(1+\frac{\epsilon_\rho}{2}+\frac{\epsilon_z}{3}\Big); \qquad b_{\rm m} = R\Big(1-\frac{\epsilon_\rho}{2}+\frac{\epsilon_z}{3}\Big); \qquad c_{\rm m} = R\Big(1-\frac{2\epsilon_z}{3}\Big),
 \label{eq:abc}
\end{equation}
where $\epsilon_\rho$ and $\epsilon_z$ are the equatorial and the polar flattenings, in principle unknown and that we want to obtain. Eqs. (\ref{eq:abc}) are equivalent, at the first order in flattenings, to those used in Eq. (\ref{eq:eprho-epz}) and are valid, at the first order in flattenings, for any triaxial ellipsoid (they are just consequences of the definitions used).

The equation of surface of this homogeneous triaxial ellipsoid in $\mathcal{F}_\mathcal{B}$ is:
\begin{equation}
 1 = \frac{X^2}{a_{\rm m}^2}+\frac{Y^2}{b_{\rm m}^2}+\frac{Z^2}{c_{\rm m}^2}.
\end{equation}

If we use the semiaxes given by Eq. (\ref{eq:abc}), the spherical coordinates
\begin{equation}
 X=\rho \sin{\widehat{\theta}} \cos{\widehat{\varphi}}; \qquad Y=\rho \sin{\widehat{\theta}} \sin{\widehat{\varphi}}; \qquad Z=\rho \cos{\widehat{\theta}}.
\end{equation}
and expanding to the first order in the flattenings, we obtain
\begin{equation}
 \rho(\widehat{\theta},\widehat{\varphi})=R\Big(1+\frac{\epsilon_\rho}{2}\sin^2{\widehat{\theta}}\cos{2\widehat{\varphi}}+\epsilon_z\Big(\frac{1}{3}-\cos^2{\widehat{\theta}}\Big)\Big).
 \label{eq:drho}
\end{equation}
\begin{figure}[h]
\begin{center}
\includegraphics[scale=0.5]{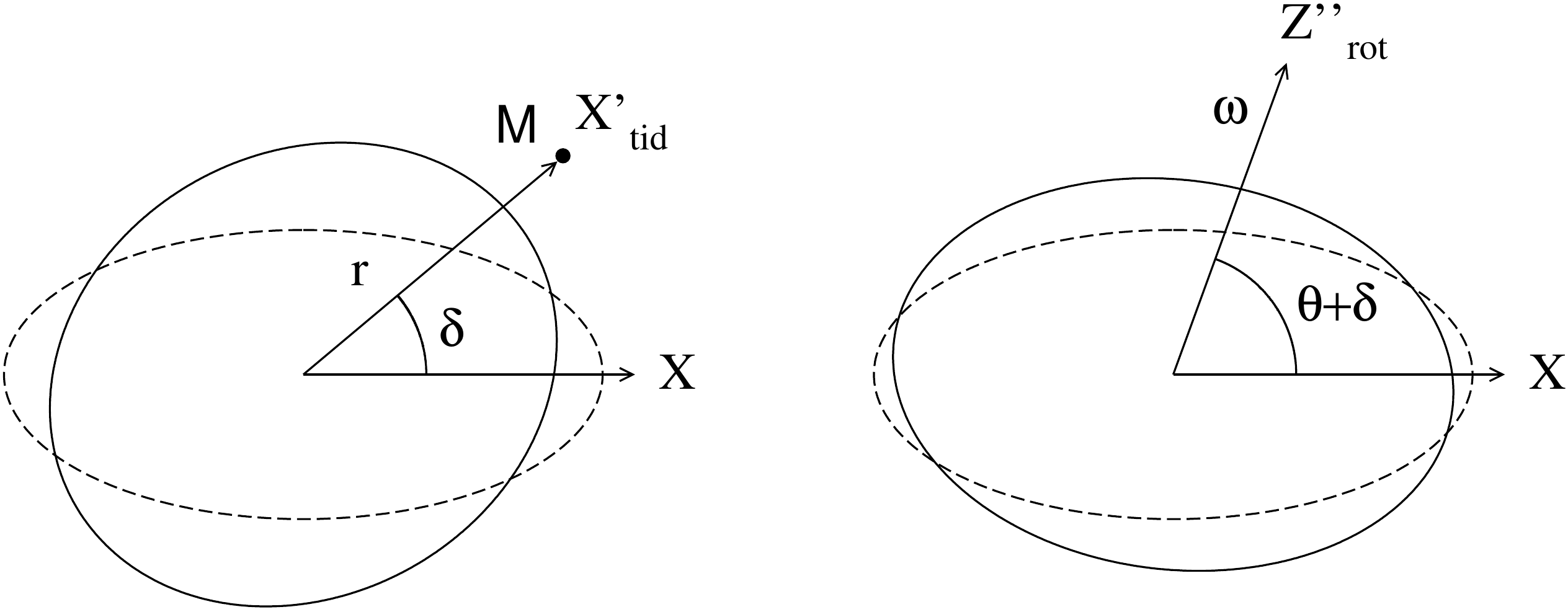}
\caption{Polar section of the equilibrium spheroid corresponding to the tide generated by \tens{M} on \tens{m} (\textit{left}) and corresponding to the rotation (\textit{right}).}
\label{fig:apB}
\end{center}
\end{figure}

For the case of the tidal deformation $\delta\rho_{\rm tid}$, we use the Jeans ellipsoid (\textit{left} panel of Fig. \ref{fig:apB}), where the equatorial and the polar flattenings are:
\begin{equation}
\epsilon_\rho^{(\rm tid)} = \epsilon_J; \qquad \epsilon_z^{(\rm tid)} = \frac{\epsilon_J}{2}.
\end{equation}
Hence, the semiaxes of this ellipsoid can be written as
\begin{equation}
 a_{\rm tid} = R\Big(1+\frac{2\epsilon_J}{3}\Big); \qquad b_{\rm tid} = c_{\rm tid} = R\Big(1-\frac{\epsilon_J}{3}\Big).
\end{equation}

The equation of surface of this homogeneous ellipsoid, in a reference frame $\mathcal{F}_r$ where the semiaxes $a_{\rm tid}$, $b_{\rm tid}$ and $c_{\rm tid}$ are aligned to the coordinates axes $X'_{\rm tid}$, $Y'_{\rm tid}$ and $Z'_{\rm tid}$, respectively, is:
\begin{equation}
 1 = \frac{X_{\rm tid}^{'2}}{a_{\rm tid}^2}+\frac{Y_{\rm tid}^{'2}}{b_{\rm tid}^2}+\frac{Z_{\rm tid}^{'2}}{c_{\rm tid}^2}.
\end{equation}
It is important to note that $X'_{\rm tid}$ is pointing toward the companion.

If $(X_{\rm tid},Y_{\rm tid},Z_{\rm tid})$, defining by
\begin{equation}
 X_{\rm tid}=\rho_{\rm tid} \sin{\widehat{\theta}} \cos{\widehat{\varphi}}; \qquad Y_{\rm tid}=\rho_{\rm tid} \sin{\widehat{\theta}} \sin{\widehat{\varphi}}; \qquad Z_{\rm tid}=\rho_{\rm tid} \cos{\widehat{\theta}},
\end{equation}
are the coordinates of the point $(X'_{\rm tid},Y'_{\rm tid},Z'_{\rm tid})$ in $\mathcal{F}_\mathcal{B}$, thus, using the rotation matrix around the second axis $\tens{R}_2(\delta)$, we obtain
\begin{eqnarray}
 X'_{\rm tid} &=& X_{\rm tid}\cos{\delta}+Z_{\rm tid}\sin{\delta}\nonumber\\
 Y'_{\rm tid} &=& Y_{\rm tid} \nonumber\\
 Z'_{\rm tid} &=&-X_{\rm tid}\sin{\delta}+Z_{\rm tid}\cos{\delta}.
\end{eqnarray}

Expanding to the first order in the flattenings, we obtain
\begin{eqnarray}
 \rho_{\rm tid}(\widehat{\theta},\widehat{\varphi})&=&R\Big(1+\frac{\epsilon_J}{2}\frac{1+\cos{2\delta}}{2}\sin^2{\widehat{\theta}}\cos{2\widehat{\varphi}}-\frac{3\epsilon_J}{4}\Big(\frac{1}{3}-\cos{2\delta}\Big)\Big(\frac{1}{3}-\cos^2{\widehat{\theta}}\Big)\nonumber\\
 &&\qquad+\frac{\epsilon_J}{4}\sin{2\delta}\sin{2\widehat{\theta}}\cos{\widehat{\varphi}}\Big)
 \label{eq:drho-tid}
\end{eqnarray}

For the case of the rotational deformation $\delta\rho_{\rm rot}$, we use the Maclaurin ellipsoid (\textit{right} panel of Fig. \ref{fig:apB}), where the equatorial and the polar flattenings are:
\begin{equation}
\epsilon_\rho^{(\rm rot)} = 0; \qquad \epsilon_z^{(\rm rot)} = \epsilon_M,
\end{equation}
thus, the semiaxes of this ellipsoid are
\begin{equation}
 a_{\rm rot} = b_{\rm rot} = R\Big(1+\frac{\epsilon_M}{3}\Big); \qquad   c_{\rm rot} = R\Big(1-\frac{2\epsilon_M}{3}\Big).
\end{equation}

The equation of surface of Maclaurin ellipsoid, in a reference frame $\mathcal{F}_\omega$ where the semiaxes $a_{\rm rot}$, $b_{\rm rot}$ and $c_{\rm rot}$ are aligned to the coordinates axes $X''_{\rm rot}$, $Y''_{\rm rot}$ and $Z''_{\rm rot}$, respectively, is:
\begin{equation}
 1 = \frac{X_{\rm rot}^{''2}}{a_{\rm rot}^2}+\frac{Y_{\rm rot}^{''2}}{b_{\rm rot}^2}+\frac{Z_{\rm rot}^{''2}}{c_{\rm rot}^2}.
\end{equation}
In this case, $Z''_{\rm rot}$ is parallel to the spin rate vector $\vec{\omega}$.

If $(X_{\rm rot},Y_{\rm rot},Z_{\rm rot})$, defining by
\begin{equation}
 X_{\rm rot}=\rho_{\rm rot} \sin{\widehat{\theta}} \cos{\widehat{\varphi}}; \qquad Y_{\rm rot}=\rho_{\rm rot} \sin{\widehat{\theta}} \sin{\widehat{\varphi}}; \qquad Z_{\rm rot}=\rho_{\rm rot} \cos{\widehat{\theta}},
\end{equation}
are the coordinates of the point $(X''_{\rm rot},Y''_{\rm rot},Z''_{\rm rot})$ in $\mathcal{F}_\mathcal{B}$, thus, using the rotation matrix $\tens{R}_2(\theta+\delta-\pi/2)$, we obtain
\begin{eqnarray}
 X''_{\rm rot} &=& X_{\rm rot}\sin{\theta_\mathcal{B}}-Z_{\rm rot}\cos{\theta_\mathcal{B}}\nonumber\\
 Y''_{\rm rot} &=& Y_{\rm rot} \nonumber\\
 Z''_{\rm rot} &=& X_{\rm rot}\cos{\theta_\mathcal{B}}+Z_{\rm rot}\sin{\theta_\mathcal{B}},
\end{eqnarray}
where $\theta_\mathcal{B}=\theta+\delta$.

Expanding to the first order in the flattenings, we obtain
\begin{eqnarray}
 \rho_{\rm rot}(\widehat{\theta},\widehat{\varphi})&=&R\Big(1-\frac{\epsilon_M}{2}\frac{1+\cos{2\theta_\mathcal{B}}}{2}\sin^2{\widehat{\theta}}\cos{2\widehat{\varphi}}+\frac{3\epsilon_M}{4}\Big(\frac{1}{3}-\cos{2\theta_\mathcal{B}}\Big)\Big(\frac{1}{3}-\cos^2{\widehat{\theta}}\Big)\nonumber\\
 &&\qquad-\frac{\epsilon_M}{4}\sin{2\theta_\mathcal{B}}\sin{2\widehat{\theta}}\cos{\widehat{\varphi}}\Big).
 \label{eq:drho-rot}
\end{eqnarray}

Finally, using Eqs. (\ref{eq:drho-def}), (\ref{eq:drho}), (\ref{eq:drho-tid}) and (\ref{eq:drho-rot}), by identification of the terms with same trigonometric arguments we obtain three equations for the flattenings $\epsilon_\rho$ and $\epsilon_z$ and the angle $\delta$:
\begin{eqnarray}
 \epsilon_\rho &=& \frac{\epsilon_J-\epsilon_M}{2}+\frac{1}{2}\Big(\epsilon_J\cos{2\delta}-\epsilon_M\cos{2\theta_\mathcal{B}}\Big)\nonumber\\
 \epsilon_z &=& \frac{\epsilon_M-\epsilon_J}{4}+\frac{3}{4}\Big(\epsilon_J\cos{2\delta}-\epsilon_M\cos{2\theta_\mathcal{B}}\Big)\nonumber\\\nonumber\\
 0 &=& \epsilon_J\sin{2\delta}-\epsilon_M\sin{2\theta_\mathcal{B}}.
 \label{eq:sys-ellip}
\end{eqnarray}

The system of equations (\ref{eq:sys-ellip}) is the same as that found in Sect. \ref{sec3} (Eq. \ref{eq:achata2}). Hence, the solution for the equatorial flattening for the polar flattening and for the angle of orientation is the same
\begin{eqnarray}
 \epsilon_\rho &=& \frac{\epsilon_J-\epsilon_M}{2}+\frac{1}{2}\sqrt{\epsilon_J^2+\epsilon_M^2-2\epsilon_J\epsilon_M\cos{2\theta}}\nonumber\\
 \epsilon_z &=& \frac{\epsilon_M-\epsilon_J}{4}+\frac{3}{4}\sqrt{\epsilon_J^2+\epsilon_M^2-2\epsilon_J\epsilon_M\cos{2\theta}},
\end{eqnarray}
and
\begin{equation}
\cos{2\delta}=\frac{\epsilon_J-\epsilon_M\cos{2\theta}}{\sqrt{\epsilon_J^2+\epsilon_M^2-2\epsilon_J\epsilon_M\cos{2\theta}}};\qquad \sin{2\delta}=\frac{\epsilon_M\sin{2\theta}}{\sqrt{\epsilon_J^2+\epsilon_M^2-2\epsilon_J\epsilon_M\cos{2\theta}}}.
\end{equation}

\section*{Appendix 3: The $\tens{B}$-matrices}{\label{ap3}}

In this section, we show the derivation of the formulas of $\tens{B}_J$ and $\tens{B}_M$ matrices presented by Eq. (\ref{eq:BM-BJ}), in Section \ref{sec4}. These matrices represent the non-dimensional form of the moment of quadrupole matrix (see Ragazzo and Ruiz, 2017) due to the rotational and the tidal deformation.

For this reason, we proceed by presenting the expressions for the principal moments of inertia $A<B<C$. From Eq. (\ref{eq:eprho-epz}), the principal moments of inertia can be expressed in terms of the equatorial and the polar flattenings $\epsilon_\rho$ and $\epsilon_z$, respectively, and of the mean body radius $R=\sqrt[3]{a_{\rm m}b_{\rm m}c_{\rm m}}$ of a general triaxial ellipsoid with semiaxes $a_{\rm m}<b_{\rm m}<c_{\rm m}$ as:
\begin{eqnarray}
 A &=& \frac{1}{5}m(b_{\rm m}^2+c_{\rm m}^2) \approx I_0 \left(1-\frac{\epsilon_\rho}{2}-\frac{\epsilon_z}{3}\right) \nonumber\\
 B &=& \frac{1}{5}m(a_{\rm m}^2+c_{\rm m}^2) \approx I_0 \left(1+\frac{\epsilon_\rho}{2}-\frac{\epsilon_z}{3}\right)\nonumber\\
 C &=& \frac{1}{5}m(a_{\rm m}^2+b_{\rm m}^2) \approx I_0 \left(1+\frac{2\epsilon_z}{3}\right),
 \label{eq:ABC}
\end{eqnarray}
where $I_0=\frac{2}{5}mR^2$. It is important to note that the rightmost expression in Eq. (\ref{eq:ABC}) is a linearization, so despite being valid for a general triaxial ellipsoid, it is only valid for ellipsoids with small flattening.

In the case in which only the deformation due to the tides is considered, the primary takes the shape of a Jeans ellipsoid (\textit{left} panel of Fig. \ref{fig:apB}), where its equatorial vertex is pointing toward the companion and its equatorial and polar flattenings are:
\begin{equation}
\epsilon_\rho^{(\rm tid)} = \epsilon_J; \qquad \epsilon_z^{(\rm tid)} = \frac{\epsilon_J}{2}.
\label{eq:achatas-rz-tid}
\end{equation}

Thus, replacing Eq. (\ref{eq:achatas-rz-tid}) in Eq. (\ref{eq:ABC}), we obtain the principal moments of inertia of the Jeans ellipsoid:
\begin{equation}
 A_{\rm tid} = R\Big(1-\frac{2\epsilon_J}{3}\Big); \qquad B_{\rm tid} = C_{\rm tid} = R\Big(1+\frac{\epsilon_J}{3}\Big).
\end{equation}

Hence, the inertia tensor, in a reference frame $\mathcal{F}_r$ where the semiaxes $a_{\rm tid}$, $b_{\rm tid}$ and $c_{\rm tid}$ are aligned to the coordinates axes $X'_{\rm tid}$, $Y'_{\rm tid}$ and $Z'_{\rm tid}$ (defined in the above appendix), respectively, is
\begin{equation}
 \tens{I}_{\mathcal{F}_r}^{\rm tid} = I_0(\mathbb{I}-\tens{B}_{\mathcal{F}_\omega}^{(J)}),
\end{equation}
where the $\tens{B}_{\mathcal{F}_r}^{(J)}$ matrix in reference frame $\mathcal{F}_r$ is:
\begin{eqnarray}
\tens{B}_{\mathcal{F}_r}^{(J)}&=&\frac{\epsilon_J}{r^2}\left(\begin{array}{ccc}
   \frac{2}{3}r^2 & 0              & 0 \\
                0 &-\frac{1}{3}r^2 & 0 \\
                0 &              0 & -\frac{1}{3}r^2  \end{array} \right).
\end{eqnarray}

On the other hand, it is important to note that the radius-vector $\vec{r}_{\mathcal{F}_r}$ in reference frame $\mathcal{F}_r$ has the simple form:
\begin{eqnarray}
\vec{r}_{\mathcal{F}_r}=\left(\begin{array}{c}
 r \\
 0 \\
 0  \end{array} \right).
\end{eqnarray}

In order to describe the matrix $\tens{B}_J$ in a general inertial reference frame $\mathcal{F}_0$, we define the rotation matrix $\tens{S}$ as
\begin{eqnarray}
 \tens{S}&=&\tens{R}_3(\varphi_0)\tens{R}_2(\theta_0-\pi/2)=\left(\begin{array}{ccc}
 \sin{\theta_0}\cos{\varphi_0} & -\sin{\varphi_0} &-\cos{\theta_0}\cos{\varphi_0} \\
 \sin{\theta_0}\sin{\varphi_0} &  \cos{\varphi_0} &-\cos{\theta_0}\sin{\varphi_0} \\
 \cos{\theta_0}              &              0 & \sin{\theta_0}\end{array} \right),
\end{eqnarray}
where $\theta_0,\varphi_0$ are the co-latitude and the longitude of the companion with respect to $\mathcal{F}_0$.

$\tens{R}_2$ and $\tens{R}_3$ represent the rotation matrices around the second and third axis, respectively:
\begin{equation}
 \tens{R}_2(\alpha) = \left(\begin{array}{ccc}
 \cos{\alpha} & 0 & \sin{\alpha} \\
            0 & 1 & 0 \\
-\sin{\alpha} & 0 & \cos{\alpha} \end{array} \right); \qquad
 \tens{R}_3(\alpha) = \left(\begin{array}{ccc}
            \cos{\alpha} & -\sin{\alpha} & 0 \\
            \sin{\alpha} & \cos{\alpha}  & 0 \\
            0 & 0 & 1\end{array} \right).
\end{equation}

Therefore, inertia tensor and the radius-vector in ${\mathcal{F}_0}$ are
\begin{eqnarray}
\tens{I}_{\mathcal{F}_0}^{\rm (tid)}\defeq \tens{I}_{\rm tid} &=& \tens{S}\tens{I}_{\mathcal{F}_r}^{\rm (tid)}\tens{S}^{-1}\nonumber\\
 &=& I_0\big(\mathbb{I}-\tens{S}\tens{B}_{\mathcal{F}_r}^{(J)}\tens{S}^{-1}\big),
\end{eqnarray}
and
\begin{eqnarray}
\vec{r}_{\mathcal{F}_0}\defeq \vec{r} &=& \tens{S}\vec{r}_{\mathcal{F}_r}=\left(\begin{array}{ccc}
   r\sin{\theta_0}\cos{\varphi_0} \\
   r\sin{\theta_0}\sin{\varphi_0} \\
   r\cos{\theta_0} \end{array} \right) =
\left(\begin{array}{c}
 x \\
 y \\
 z  \end{array} \right),
 \end{eqnarray}
respectively.

Therefore, the matrix $\tens{S}\tens{B}_{\mathcal{F}_r}^{(J)}\tens{S}^{-1}$ is $\tens{B}_{\mathcal{F}_0}^{(J)}\defeq\tens{B}_J$ in reference frame $\mathcal{F}_0$:
\begin{eqnarray}
\tens{B}_J &=& \tens{S}\tens{B}_{\mathcal{F}_r}^{(J)}\tens{S}^{-1}\nonumber\\
&=&\frac{\epsilon_J}{r^2}\left(\begin{array}{ccc}
  r^2 \sin^2{\theta_0} \cos^2{\varphi_0}-\frac{1}{3}r^2 & r^2\sin^2{\theta_0}\sin{\varphi_0}\cos{\varphi_0} &  r^2\cos{\theta_0}\sin{\theta_0}\cos{\varphi_0} \\
r^2\sin^2{\theta_0}\sin{\varphi_0}\cos{\varphi_0} & r^2 \sin^2{\theta_0} \sin^2{\varphi_0}-\frac{1}{3}r^2 & r^2\cos{\theta_0}\sin{\theta_0}\sin{\varphi_0} \\
r^2\cos{\theta_0}\sin{\theta_0}\cos{\varphi_0}  &  r^2\cos{\theta_0}\sin{\theta_0}\sin{\varphi_0}           & r^2 \cos^2{\theta_0}-\frac{1}{3}r^2 \end{array} \right)\nonumber\\
&=&\frac{\epsilon_J}{r^2}\left(\begin{array}{ccc}
   x^2-\frac{1}{3}r^2 & xy &  xz \\
yx & y^2-\frac{1}{3}r^2  & yz \\
zx  & zy        & z^2-\frac{1}{3}r^2 \end{array} \right) = \frac{\epsilon_J}{r^2}\Big(\vec{r}\otimes\vec{r}-\frac{r^2}{3}\mathbb{I} \Big).
\end{eqnarray}

In the case in which only the deformation due to the rotation is considered, the primary takes the shape of a Maclaurin ellipsoid (\textit{right} panel of Fig. \ref{fig:apB}), where its axial axis of symmetry is parallel to the angular velocity vector and its equatorial and polar flattenings are:
\begin{equation}
\epsilon_\rho^{(\rm rot)} = 0; \qquad \epsilon_z^{(\rm rot)} = \epsilon_M.
\label{eq:achatas-rz-rot}
\end{equation}

Thus, replacing Eq. (\ref{eq:achatas-rz-rot}) in Eq. (\ref{eq:ABC}), we obtain the principal moments of inertia of the Maclaurin ellipsoid:
\begin{equation}
 A_{\rm rot} = B_{\rm rot} = I_0\Big(1-\frac{\epsilon_M}{3}\Big); \qquad   C_{\rm rot} = I_0\Big(1+\frac{2\epsilon_M}{3}\Big).
\end{equation}

Hence, the inertia tensor, in a reference frame $\mathcal{F}_\omega$ where the semiaxes $a_{\rm rot}$, $b_{\rm rot}$ and $c_{\rm rot}$ are aligned to the coordinates axes $X''_{\rm rot}$, $Y''_{\rm rot}$ and $Z''_{\rm rot}$ (defined in the above appendix), respectively, is
\begin{equation}
 \tens{I}_{\mathcal{F}_\omega}^{\rm (rot)} = I_0(\mathbb{I}-\tens{B}_{\mathcal{F}_\omega}^{(M)})
\end{equation}
where the $\tens{B}_{\mathcal{F}_\omega}^{(M)}$ matrix in reference frame $\mathcal{F}_\omega$ is:
\begin{eqnarray}
\tens{B}_{\mathcal{F}_\omega}^{(M)}&=&\frac{\epsilon_M}{\omega^2}\left(\begin{array}{ccc}
    \frac{1}{3}\omega^2 & 0                   & 0 \\
                      0 & \frac{1}{3}\omega^2 & 0 \\
                      0 &                   0 & -\frac{2}{3}\omega^2  \end{array} \right).
\end{eqnarray}

On the other hand, it is important to note that the angular velocity vector $\vec{\omega}_{\mathcal{F}_\omega}$ in reference frame $\mathcal{F}_\omega$ has the simple form:
\begin{eqnarray}
\vec{\omega}_{\mathcal{F}_\omega}=\left(\begin{array}{c}
 0 \\
 0 \\
 \omega  \end{array} \right).
\end{eqnarray}

In order to describe the matrix $\tens{B}_M$ in a general inertial reference frame $\mathcal{F}_\omega$, we define the rotation matrix $\tens{R}$ as
\begin{eqnarray}
\tens{R}&=&\tens{R}_3(\psi)\tens{R}_1(J) =\left(\begin{array}{ccc}
    \cos{\psi} & -\cos{J}\sin{\psi} &  \sin{J}\sin{\psi} \\
\sin{\psi} &  \cos{J}\cos{\psi} & -\sin{J}\cos{\psi} \\
        0  &  \sin{J}           &  \cos{J}           \end{array} \right),
\end{eqnarray}
where $J$ is the inclination of the equatorial plane of the primary (defined as the plane perpendicular to the angular velocity vector) with respect to the $x_0-y_0$ inertial plane, and $\psi$ is the longitude of the ascending node of the rotational equator.

$\tens{R}_1$ and $\tens{R}_3$ represent the rotation matrices around the first and third axis, respectively:
\begin{equation}
 \tens{R}_1(\alpha) = \left(\begin{array}{ccc}
            1 & 0 & 0 \\
            0 & \cos{\alpha} & -\sin{\alpha} \\
            0 & \sin{\alpha} & \cos{\alpha} \end{array} \right).
\end{equation}

Therefore, inertia tensor and the angular velocity vector in ${\mathcal{F}_0}$ are
\begin{eqnarray}
 \tens{I}_{\mathcal{F}_0}^{\rm (rot)}\defeq \tens{I}_{\rm rot} &=& \tens{R}\tens{I}_{\mathcal{F}_\omega}^{\rm (rot)}\tens{R}^{-1}\nonumber\\
 &=& I_0\big(\mathbb{I}-\tens{R}\tens{B}_{\mathcal{F}_\omega}^{(M)}\tens{R}^{-1}\big),
\end{eqnarray}
and
\begin{eqnarray}
 \vec{\omega}_{\mathcal{F}_0}\defeq\vec{\omega} &=& \tens{R}\vec{\omega}_{\mathcal{F}_\omega}=\left(\begin{array}{ccc}
   \omega\sin{J}\sin{\psi} \\
  -\omega\sin{J}\cos{\psi} \\
   \omega\cos{J}  \end{array} \right) =
\left(\begin{array}{c}
 \omega_x \\
 \omega_y \\
 \omega_z  \end{array} \right),
 \end{eqnarray}
respectively.

Therefore, the matrix $\tens{R}\tens{B}_{\mathcal{F}_\omega}^{(M)}\tens{R}^{-1}$ is $\tens{B}_{\mathcal{F}_0}^{(M)}\defeq\tens{B}_M$ in reference frame $\mathcal{F}_0$:
\begin{eqnarray}
\tens{B}_M &=& \tens{R}\tens{B}_{\mathcal{F}_\omega}^{(M)}\tens{R}^{-1}\nonumber\\
&=&\frac{\epsilon_M}{\omega^2}\left(\begin{array}{ccc}
  \frac{1}{3}\omega^2 -\omega^2 \sin^2{J} \sin^2{\psi} & \omega^2\sin^2{J}\sin{\psi}\cos{\psi} &  -\omega^2\cos{J}\sin{J}\sin{\psi} \\
\omega^2\sin^2{J}\sin{\psi}\cos{\psi} & \frac{1}{3}\omega^2 -\omega^2 \sin^2{J} \cos^2{\psi} & \omega^2\cos{J}\sin{J}\cos{\psi} \\
-\omega^2\cos{J}\sin{J}\sin{\psi}  &  \omega^2\cos{J}\sin{J}\cos{\psi}           & \frac{1}{3}\omega^2 -\omega^2 \cos^2{J} \end{array} \right)\nonumber\\
&=&\frac{\epsilon_M}{\omega^2}\left(\begin{array}{ccc}
  \frac{1}{3}\omega^2 -\omega_x^2 & -\omega_x\omega_y &  -\omega_x\omega_z \\
-\omega_y\omega_x & \frac{1}{3}\omega^2 -\omega_y^2 & -\omega_y\omega_z \\
-\omega_z\omega_x  & -\omega_z\omega_y        & \frac{1}{3}\omega^2 -\omega_z^2 \end{array} \right) = \frac{\epsilon_M}{\omega^2}\Big(\frac{\omega^2}{3}\mathbb{I} - \vec{\omega}\otimes\vec{\omega}\Big).
\end{eqnarray}

Finally, and as a consequence of the linearity of the deformations shown in the previous section (that is, when these deformations are small enough for the linear expansions in the flattening to be valid), the inertial tensor of a body deformed by the action of the tides and the rotation in the inertial reference frame $\mathcal{F}_0$ can be written as the sum of the two components presented in this section:
\begin{equation}
  \tens{I}_{\mathcal{F}_0}\defeq \tens{I} = I_0(\mathbb{I}-\tens{B}_J-\tens{B}_M).
\end{equation}

%
%
%

\end{document}